\documentclass[twocolumn]{aastex62}

\usepackage{comment}
\usepackage{bm}
\usepackage{float}

\graphicspath{{./}{figures/}}

\shorttitle{Proximity zone sizes of faint quasars at $z\sim6$}
\shortauthors{Ishimoto et al.}

\begin{document}
\bibliographystyle{aasjournal}
\turnoffeditone

\title{Subaru High-$z$ Exploration of Low-Luminosity Quasars (SHELLQs). XI.
Proximity Zone Analysis for Faint Quasar Spectra at $z\sim 6$}

\correspondingauthor{Rikako Ishimoto}
\email{ishimoto@astron.s.u-tokyo.ac.jp}

\author[0000-0002-2134-2902]{Rikako Ishimoto}
\affiliation{Department of Astronomy, Graduate School of Science, The University of Tokyo, 7-3-1 Hongo, Bunkyo, Tokyo 113-0033, Japan}

\author[0000-0001-5493-6259]{Nobunari Kashikawa}
\affiliation{Department of Astronomy, Graduate School of Science, The University of Tokyo, 7-3-1 Hongo, Bunkyo, Tokyo 113-0033, Japan}

\author[0000-0003-2984-6803]{Masafusa Onoue}
\affiliation{Max Planck Institut f\"{u}r Astronomie, K\"{o}nigstuhl 17, D-69117 Heidelberg, Germany}

\author{Yoshiki Matsuoka}
\affiliation{Research Center for Space and Cosmic Evolution, Ehime University, Matsuyama, Ehime 790-8577, Japan.}

\author[0000-0001-9452-0813]{Takuma Izumi}
\affiliation{National Astronomical Observatory of Japan, 2-21-1 Osawa, Mitaka, Tokyo 181-8588, Japan}
\affiliation{Department of Astronomical Science, Graduate University for Advanced Studies (SOKENDAI), Mitaka, Tokyo 181-8588, Japan.}

\author[0000-0002-0106-7755]{Michael A. Strauss}
\affil{Princeton University Observatory, Peyton Hall, Princeton, NJ 08544, USA.}

\author[0000-0001-7201-5066]{Seiji Fujimoto}
\affil{Cosmic DAWN Center, Copenhagen, Denmark}
\affil{Niels Bohr Institute, University of Copenhagen, Lyngbyvej 2, DK-2100, Copenhagen, Denmark}

\author[0000-0001-6186-8792]{Masatoshi Imanishi}
\affil{National Astronomical Observatory of Japan, 2-21-1 Osawa, Mitaka, Tokyo 181-8588, Japan}
\affil{Department of Astronomical Science, Graduate University for Advanced Studies (SOKENDAI), Mitaka, Tokyo 181-8588, Japan.}

\author[0000-0002-9453-0381]{Kei Ito}
\affil{National Astronomical Observatory of Japan, 2-21-1 Osawa, Mitaka, Tokyo 181-8588, Japan}
\affil{Department of Astronomical Science, Graduate University for Advanced Studies (SOKENDAI), Mitaka, Tokyo 181-8588, Japan.}

\author[0000-0002-4923-3281]{Kazushi Iwasawa}
\affil{ICREA and Institut de Ci{\`e}ncies del Cosmos, Universitat de Barcelona, IEEC-UB, Mart{\'i} i Franqu{\`e}s, 1, 08028 Barcelona, Spain.}

\author{Toshihiro Kawaguchi}
\affil{Department of Economics, Management and Information Science, Onomichi City University, Onomichi, Hiroshima 722-8506, Japan.}

\author[0000-0003-1700-5740]{Chien-Hsiu Lee}
\affil{National Optical Astronomy Observatory, 950 North Cherry Avenue, Tucson, AZ 85719, USA.}

\author[0000-0002-2725-302X]{Yongming Liang}
\affil{National Astronomical Observatory of Japan, 2-21-1 Osawa, Mitaka, Tokyo 181-8588, Japan}
\affil{Department of Astronomical Science, Graduate University for Advanced Studies (SOKENDAI), Mitaka, Tokyo 181-8588, Japan.}

\author[0000-0002-4965-6524]{Ting-Yi Lu}
\affil{Institute of Astronomy and Department of Physics, National Tsing Hua University, Hsinchu 30013, Taiwan.}

\author[0000-0002-8857-2905]{Rieko Momose}
\affiliation{Department of Astronomy, Graduate School of Science, The University of Tokyo, 7-3-1 Hongo, Bunkyo, Tokyo 113-0033, Japan}

\author[0000-0002-3531-7863]{Yoshiki Toba}
\affil{Department of Astronomy, Kyoto University, Sakyo-ku, Kyoto, Kyoto 606-8502, Japan.}
\affil{Institute of Astronomy and Astrophysics, Academia Sinica, Taipei, 10617, Taiwan.}
\affiliation{Research Center for Space and Cosmic Evolution, Ehime University, Matsuyama, Ehime 790-8577, Japan.}

\author{Hisakazu Uchiyama}
\affil{National Astronomical Observatory of Japan, 2-21-1 Osawa, Mitaka, Tokyo 181-8588, Japan}

\begin{abstract}
We present measurements of the size of the quasar proximity zone ($R_p$)
for \edit1{eleven} low-luminosity ($-26.16\leq M_{1450}\leq-22.83$) quasars at $z\sim6$, discovered by the Subaru High-$z$ Exploration of Low-Luminosity Quasars project (SHELLQs).
Our faint quasar sample expands the $R_p$ measurement down to $M_{1450}=-22.83$ mag, where more common quasar populations dominate at the epoch.
We restrict the sample to quasars whose systemic redshifts have been precisely measured by [\ion{C}{2}] 158 \micron\ or \ion{Mg}{2} $\lambda$2798 emission lines.
\edit1{We also update the $R_p$ measurements for 26 luminous quasars presented in \citet{Eilers2017} by using the latest systemic redshift results.}
\deleted{We \edit1{augment} our sample size by measuring \deleted{the} $R_p$ for 26 luminous quasars presented in \citet{Eilers2017} with updated redshift measurements.}
The luminosity dependence on $R_p$ is found to be consistent with the theoretical prediction assuming highly ionized intergalactic medium.
We \edit1{find} a shallow redshift evolution of the luminosity-corrected $R_p$, \edit1{$R_{p,{\rm corr}}^{-25}$ ($R_{p, \rm corr}^{-25}\propto(1+z)^{-3.79\pm1.72}$)} over $5.8\lesssim z \lesssim6.6$.
This trend is \edit1{steeper than that of \citet{Eilers2017}, but significantly shallower than those of the earlier studies.}
Our results suggest that $R_{p,\rm corr}$ is insensitive to the neutral fraction of the universe at $z\sim6$.
\edit1{Four} quasars show exceptionally small \edit1{$R_{p,\rm corr}^{-25}$ ($ \lesssim0.90$  proper Mpc)},
which could be the result of their young age ($<10^4$ yr) in the reionization epoch, though statistics is still small.

\end{abstract}


\section{Introduction}\label{sec:intro}
Cosmic reionization  was a  key event in the early universe. 
After recombination at $z\sim1100$,  the neutral intergalactic medium (IGM) was ionized by  ultraviolet radiation from the first  generations of stars and galaxies.
 Recent observations of the polarization of the cosmic microwave background (CMB)  imply a reionization redshift $z_{\rm{reion}} = 7.7\pm0.7$
\citep{PlanckCollaboration2018} assuming instantaneous reionization,
but when and how reionization proceeded is still under debate.

 In recent years, the number of known quasars in the early universe at $z> 6$ has increased dramatically \citep{Reed2017,Banados2018,Matsuoka2019,Wang2019}.
High-$z$ ($z\gtrsim6$) quasar spectra are used as a powerful probe of the state of the IGM in the reionization era.
Absorption by   neutral hydrogen in the Ly$\alpha$ forest in the IGM yields some characteristic features in quasar spectra.
 The observation of Ly$\alpha$ optical depth,
has revealed  a steep increase  in the IGM neutral fraction, $f_{\rm HI}$, and its scatter at $z>5.5$ \citep{Fan2005,Becker2015,Eilers2018a}.
However,  the Ly$\alpha$ forest  cannot be used to measure $f_{\rm HI}$ at $z\gtrsim6$ because it saturates when the IGM neutral fraction is high ($f_{\rm{HI}}>10^{-4}$) \citep{Gunn1965}.

\edit1{There is another approach to measure $f_{\rm HI}$,} the size of the proximity zone around quasars.
 The proximity  zone is an ionized region around a quasar generated by intense quasar radiation \citep[e.g.,][]{Cen2000}.
It  has been argued that proximity zone radius, $R_p$, evolves with redshift as a good proxy of neutral fraction.
 If the IGM is uniform and the quasar lifetime is much less than  both the hydrogen recombination time and the age of the universe at  that redshift,
 $R_p$ is  given by
\begin{eqnarray}
    R_p=f_{\rm HI}^{-1/3}\left(\frac{\dot{N}_Q}{6.5\times10^{57}\ {\rm s}^{-1}}\right)^{1/3} \nonumber \\* 
   \times \left(\frac{t_Q}{2\times10^7\ {\rm yr}}\right)^{1/3}\frac{7}{1+z_Q} \ {\rm proper\ Mpc},
    \label{eq:Rp_haiman}
\end{eqnarray}
where $\dot{N_Q}$ is  the rate of ionizing  photons emitted by the quasar
and $t_Q$ is the quasar age \citep{Haiman2002}.
Early observational studies found a steep evolution  with redshift of proximity zone sizes.
 For example, \citet{Fan2005} measured proximity zone sizes of 19 quasars at  $5.74<z<6.42$ and found that $R_p$  decreases rapidly toward higher redshifts.
\citet{Carilli2010} analyzed the proximity zone sizes of 27 quasars with more accurate redshifts and  came to the same conclusion.
\citet{Mortlock2011}, \citet{Venemans2015}, and \citet{Banados2018} extended  the measurements to $z\sim7$ quasars and confirmed  this trend.

On the other hand, \citet{Bolton2007} used hydrodynamical simulations showing that  the observed $R_p$  differs from the true radii of  the ionized region.
In  a highly ionized IGM,  the observed $R_p$ approximates the classical proximity zone, 
which is determined  solely by the quasar luminosity, and does not correspond to the extent of an \ion{H}{2} region expanding into a neutral IGM. 
This leads to  substantial underestimate of the distance to the ionizing front around quasars in the highly ionized regime.

\citet{Eilers2017}  systematically measured $R_p$  for 34 luminous quasars, and found shallower redshift evolution of  luminosity corrected $R_p$ ($\propto(1+z)^{-1.44}$) than those of  previous  studies.
The result is consistent with the prediction from hydrodynamical simulations by \citet{Bolton2007},  
suggesting that $R_p$ is insensitive to the neutral fraction of  the IGM.
 \citet{Mazzucchelli2017} also found the same shallow evolution for $z>6.5$ quasars.
They also discovered quasars having $R_p$ as small as $<1$ proper Mpc (pMpc)  after correcting by luminosity.
This result  implies that such quasars are young ($<10^5$ yr) \citep{Eilers2017,Eilers2018,Eilers2020}.
\citet{Davies2019} used their radiative transfer simulation to predict the time evolution of quasar proximity zone size and showed that these small proximity zone  size could be reproduced when the IGM gas has not yet reached photoionization equilibrium around young quasars.

However, all  these studies are based  only on luminous quasars, which might reside in  unusually overdense regions at the epoch.
 Fainter quasars are more common in the universe  and test the luminosity dependence in Equation (\ref{eq:Rp_haiman}) \citep[e.g.,][]{Matsuoka2019_d,Kulkarni2019}.
Therefore, it is important to expand the dynamic range of luminosity to the faint end.
Moreover, since Equation (\ref{eq:Rp_haiman})  assumes a radiative equilibrium between the IGM and  the quasar luminosity  at all redshifts including $z\sim6$,
observational measurements  over a wide luminosity range will give insights into the physics  of the proximity zone. 
 This study,  for the first time, measures proximity zone sizes  for faint ($-26.16\leq M_{1450}\leq-22.83$) quasars at $z\sim6$ to  explore the luminosity dependence and robustness of Equation (\ref{eq:Rp_haiman}).

In Section \ref{sec:sample}, we describe  the quasar sample  we use in this work.
We describe  our method to predict intrinsic quasar spectra and to measure $R_p$ in Section \ref{sec:method}. 
We present the dependence  of $R_p$ on quasar luminosity and redshift and discuss the results in Section \ref{sec:results}.
 We summarize our results in Section \ref{sec:summary}.

Throughout this paper, we adopt a  flat $\Lambda$CDM cosmology with  $\Omega_m=0.307$ and $H_0=67.8$  km s$^{-1}$ pMpc$^{-1}$ \citep{Planck2014}.
 
\section{Quasar Sample}\label{sec:sample}
\subsection{Faint Quasars}
Our faint sample consists of \edit1{eleven} quasars at $5.93\leq z\leq6.56$  (Table \ref{tab:sample}).
All these quasars were discovered by  the Subaru High-$z$ Exploration of Low-Luminosity Quasars project (SHELLQs) using Hyper Suprime-Cam  \edit1{\citep[HSC;][]{Miyazaki2018,Komiyama2018, Kawanomoto2018, Furusawa2018}} on the Subaru Telescope \citep[e.g.,][]{Matsuoka2016,Matsuoka2018a,Matsuoka2018}.
The spectroscopic identification was carried out with the Faint Object Camera and Spectrograph \citep[FOCAS;][]{Kashikawa2002} mounted on the Subaru Telescope,  for J0859+0022,  J1153+0055, J1202$-$0057, J1208$-$0200,  J2216$-$0116, and \edit1{J2304+0045} and the Optical System for Imaging and low-intermediate-Resolution Integrated Spectroscopy \citep[OSIRIS;][]{Cepa2000} mounted on the Gran Telescopio Canarias,  for J0921+0007, J1406$-$0116, J1545+4232, J2216$-$0116, J2228+0152, and J2239+0207.
FOCAS provides spectral coverage from $\lambda_{\rm{obs}}=0.75\ \mu$m to $1.05\ \mu$m with a resolution $R\sim 1200$,
and OSIRIS provides spectral coverage from $\lambda_{\rm{obs}}=0.74\ \mu$m to 1.0 $\mu$m with a resolution $R\sim 1500$.
The exposure times are 170 minutes for J1202$-$0057, and 15 or 30 minutes for the other quasars.

 Accurate redshift measurements needed for an accurate prediction of the intrinsic spectra, are important in measuring  $R_p$.
The redshifts of \edit1{eight} of these quasars are from \citet[][in prep.]{Izumi2018,Izumi2019} and have been accurately measured by [\ion{C}{2}] 158\ \micron\ emission line.
The three quasars,  J0921+0007, J1406$-$0116, and J1545+4232 have \ion{Mg}{2} $\lambda$2798 redshifts, as well as black hole mass ($M_{\rm BH}$) and Eddington ratio ($L_{\rm bol}/L_{\rm Edd}$) measured from K-band spectra taken by Subaru/MOIRCS (ID: S19A-015, PI: M.Onoue).
The \ion{Mg}{2} redshifts were derived from the peaks of the best-fit single Gaussian profiles of the emission lines, for which the power-law continuum and the rest-frame UV iron pseudo-continuum were subtracted beforehand with the empirical iron template of \citet{Vestergaard2001}.
More details of the observations and the spectral analysis will be described in a forthcoming paper (Onoue et al. in prep.).
One broad absorption line (BAL) quasar, J1205$-$0000, is excluded from our sample because it is difficult to determine its intrinsic spectrum.
The absolute magnitude $M_{1450}$  of each quasar is taken from \citet{Matsuoka2018a,Matsuoka2018} and \citet{Onoue2019}.
They were obtained by converting UV magnitudes, assuming the power-law continuum slope of $\alpha_\lambda = -1.5$ ($F_\lambda\propto\lambda^{\alpha_\lambda}$) for J1202$-$0057, J2228+0152, \edit1{and J2304+0045} \citep{Matsuoka2018a,Matsuoka2018}, and  by fitting $\alpha_\lambda$ for the other quasars \citep{Onoue2019}.

\begin{deluxetable*}{lLLLllLLL}[htbp]
    \tablecaption{Overview of Our Faint Sample and Proximity  Zone Sizes\label{tab:sample}}
    \tablewidth{0pt}
    \tablehead{
    \colhead{Name} &\colhead{R.A.}&\colhead{Decl.}& \colhead{$z$} &\colhead{Redshift Line}
    &\colhead{References}&\colhead{$M_{1450}$ (mag)} & \colhead{$R_p$ (pMpc)}&\colhead{$R_{p, \rm corr}^{-25}$ (pMpc)}
    }
    \startdata
   J0859+0022 & 08^{\rm{h}}59^{\rm{m}}07^{\rm{s}}.19 & +00\degr22\arcmin55\arcsec.9 & 6.3903^{+0.0005}_{-0.0005} & [\ion{C}{2}] & 1& -23.10\pm0.27&1.14\pm0.03&3.14\pm0.09\\
J0921+0007 & 09^{\rm{h}}21^{\rm{m}}20^{\rm{s}}.56 & +00\degr07\arcmin22\arcsec.9 & 6.563^{+0.002}_{-0.001} & \ion{Mg}{2} &3& -26.16\pm0.29&3.05\pm0.45&1.64\pm0.24\\
J1152+0055 & 11^{\rm{h}}52^{\rm{m}}21^{\rm{s}}.27 & +00\degr55\arcmin36\arcsec.6 & 6.3637^{+0.0005}_{-0.0005} & [\ion{C}{2}] &1& -25.08\pm0.07&2.67\pm0.03&2.56\pm0.03\\
J1202$-$0057 & 12^{\rm{h}}02^{\rm{m}}46^{\rm{s}}.37 & -00\degr57\arcmin01\arcsec.7 & 5.9289^{+0.0002}_{-0.0002} & [\ion{C}{2}] &1& -22.83\pm0.08&0.74\pm0.01&2.34\pm0.04\\
J1208$-$0200 & 12^{\rm{h}}08^{\rm{m}}59^{\rm{s}}.23 & -02\degr00\arcmin34\arcsec.8 & 6.1165^{+0.0002}_{-0.0002} & [\ion{C}{2}] &2& -24.36\pm0.09&0.62\pm0.01&0.87\pm0.02\\
J1406$-$0116 & 14^{\rm{h}}06^{\rm{m}}29^{\rm{s}}.12 & -01\degr16\arcmin11\arcsec.2 & 6.292^{+0.002}_{-0.002} & \ion{Mg}{2} &3& -24.76\pm0.18&0.14\pm0.05&0.16\pm0.05\\
J1545+4232 & 15^{\rm{h}}45^{\rm{m}}05^{\rm{s}}.62 & +42\degr32\arcmin11\arcsec.6 & 6.511^{+0.003}_{-0.004} & \ion{Mg}{2} &3& -24.76\pm0.17&2.14\pm0.18&2.43\pm0.20\\
J2216$-$0016 & 22^{\rm{h}}16^{\rm{m}}44^{\rm{s}}.47 & -00\degr16\arcmin50\arcsec.1 & 6.0962^{+0.0003}_{-0.0003}& [\ion{C}{2}] &1& -23.65\pm0.20&0.66\pm0.02&1.36\pm0.04\\
J2228+0152 & 22^{\rm{h}}28^{\rm{m}}27^{\rm{s}}.83 & +01\degr28\arcmin09\arcsec.5 & 6.0805^{+0.0004}_{-0.0004} & [\ion{C}{2}] &2& -24.00\pm0.04&2.11\pm0.02&3.60\pm0.04\\
J2239+0207 & 22^{\rm{h}}39^{\rm{m}}47^{\rm{s}}.47 & +02\degr07\arcmin47\arcsec.5 & 6.2497^{+0.0004}_{-0.0004} & [\ion{C}{2}] &2& -24.60\pm0.15&1.65\pm0.02&2.04\pm0.03\\
J2304+0045 & 23^{\rm{h}}04^{\rm{m}}22^{\rm{s}}.97 & +00\degr45\arcmin05\arcsec.4 & 6.3504^{+0.0002}_{-0.0002} & [\ion{C}{2}]&4& -24.28\pm0.03 &1.15\pm0.01&1.68\pm0.02\\
    \enddata 
    \tablecomments{The columns show the object name, coordinates, the redshift and its error, the lines used to measure redshift, absolute magnitude $M_{1450}$, proximity zone sizes $R_p$, and luminosity corrected proximity zone sizes $R_{p, \rm corr}^{-25}$.\\
    References for redshifts.(1) \citet{Izumi2018}, (2)\citet{Izumi2019}, (3) Onoue et al. (in prep), (4) Izumi et al. (in prep)}
\end{deluxetable*}

\subsection{Bright Quasars}
In addition to our new quasar spectra, we use the sample of luminous quasars analyzed in \citet{Eilers2017}. 
These spectra are taken from \texttt{igmspec}\footnote{\url{http://specdb.readthedocs.io/en/latest/igmspec.html}} database.
We exclude those quasars whose redshifts were measured based on the Ly$\alpha$ emission line alone, as this line usually gives a redshift uncertainty as large as $\sim1000$ km s$^{-1}$, in order to unify the redshift accuracy with our faint quasar sample.
We also update some redshifts which have newly measured  [\ion{C}{2}] or \ion{Mg}{2} lines \citep{Willott2017,Decarli2018,Shen2019}.
We also exclude J0100+2802, 
\edit1{because \citet{Fujimoto2019} suggested this extremely bright quasar could be amplified by gravitational lensing, while there is still debate for the interpretation.}
\footnote{Actually, if the lensing hypothesis is correct, the measured proximity zone size for J0100+2802 \citep[$R_p=7.12$ pMpc;][]{Eilers2017} is too large for the $M_{1450}=-22.51$ after correcting by the inferred magnification factor $\mu =450$ (see Figure \ref{fig:M-Rp}).
\edit1{On the other hand, it is also argued that the $R_p$ measurement is  exceptionally smaller than the prediction from its uniquely bright observed luminosity of $M_{1450} = -29.26$ \citep{Eilers2017}.}} 
In the end, we use 26 quasar spectra from \citet{Eilers2017}, the systemic redshfhits of which are determined with [\ion{C}{2}], \ion{Mg}{2}, or CO emission lines, as summarized in Table \ref{tab:Esample}.

\begin{deluxetable*}{lLLLllLLL}[ht!]
    \tablecaption{Overview of Our Bright Sample and Proximity  Zone Sizes \label{tab:Esample}}
    \tablewidth{0pt}
    \tablehead{
    \colhead{Name} & \colhead{R.A.}&\colhead{Decl.}&\colhead{$z$} &
    \colhead{Redshift Line}&References& \colhead{$M_{1450}$ (mag)}&\colhead{$R_{p}$ (pMpc)} &\colhead{$R_{p, \rm corr}^{-25}$ (pMpc)}
    }
    \startdata
    J0002+2550 &  00^{\rm{h}}02^{\rm{m}}39^{\rm{s}}.39 &  +25\degr50\arcmin34\arcsec.96 &  5.818\pm0.007\tablenotemark{*} &  \ion{Mg}{2} &15&   -27.31 & 8.83\pm0.46& 2.58\pm0.13\\
J0005$-$0006 &  00^{\rm{h}}05^{\rm{m}}52^{\rm{s}}.34 &  -00\degr06\arcmin55\arcsec.80 &  5.844\pm0.001 &  \ion{Mg}{2} & 6 & -25.73 & 2.91\pm0.06& 1.97\pm0.04\\
J0050+3445 &  00^{\rm{h}}55^{\rm{m}}02^{\rm{s}}.91 &  +34\degr45\arcmin21\arcsec.65 &  6.253\pm0.003 &  \ion{Mg}{2} & 5 & -26.70 & 3.96\pm0.17& 1.60\pm0.07\\
J0148+0600 &  01^{\rm{h}}48^{\rm{m}}37^{\rm{s}}.64 &  +06\degr00\arcmin20\arcsec.06 &  5.98\pm0.01 &  \ion{Mg}{2} & 10 & -27.39 & 6.11\pm0.64& 1.71\pm0.18\\
J0210$-$0456 &  02^{\rm{h}}10^{\rm{m}}13^{\rm{s}}.19 &  -04\degr56\arcmin20\arcsec.90 &  6.4323\pm0.0005 &  [\ion{C}{2}] & 8 &  -24.53 & 1.38\pm0.03& 1.77\pm0.04\\
J0226+0302 &  02^{\rm{h}}26^{\rm{m}}01^{\rm{s}}.87 &  +03\degr02\arcmin59\arcsec.42 &  6.5412\pm0.0018 &  [\ion{C}{2}]& 9 & -27.33 & 3.66\pm0.09& 1.06\pm0.03\\
J0227$-$0605 &  02^{\rm{h}}27^{\rm{m}}43^{\rm{s}}.29 &  -06\degr05\arcmin30\arcsec.20 &  6.212\pm0.007\tablenotemark{*} &  \ion{Mg}{2} & 15 &  -25.28 & 2.27\pm0.40& 1.95\pm0.35\\
J0303$-$0019 &  03^{\rm{h}}03^{\rm{m}}31^{\rm{s}}.40 &  -00\degr19\arcmin12\arcsec.90 &  6.078\pm0.007 &  \ion{Mg}{2}& 3 & -25.56 & 2.28\pm0.44& 1.69\pm0.33\\
J0836+0054 &  08^{\rm{h}}36^{\rm{m}}43^{\rm{s}}.86 &  +00\degr54\arcmin53\arcsec.26 &  5.810\pm0.003 &  \ion{Mg}{2}& 2 & -27.75 & 5.16\pm0.20& 1.19\pm0.05\\
J0842+1218 &  08^{\rm{h}}42^{\rm{m}}29^{\rm{s}}.43 &  +12\degr18\arcmin50\arcsec.58 &  6.0763\pm0.0005\tablenotemark{*}& [\ion{C}{2}] & 14 & -26.91 & 6.95\pm0.04& 2.52\pm0.01\\
J0927+2001 &  09^{\rm{h}}27^{\rm{m}}21^{\rm{s}}.82 &  +20\degr01\arcmin23\arcsec.64 &  5.7722\pm0.0006 &  CO & 4 &  -26.76 & 4.69\pm0.05& 1.84\pm0.02\\
J1030+0524 &  10^{\rm{h}}30^{\rm{m}}27^{\rm{s}}.11 &  +05\degr24\arcmin55\arcsec.06 &  6.309\pm0.009 &  \ion{Mg}{2} & 1 & -26.99 & 6.00\pm0.51& 2.08\pm0.18\\
J1137+3549 &  11^{\rm{h}}37^{\rm{m}}17^{\rm{s}}.73 &  +35\degr49\arcmin56\arcsec.85 &  6.009\pm0.010\tablenotemark{*} &  \ion{Mg}{2} & 15 &  -27.36 & 5.81\pm0.62& 1.66\pm0.18\\
J1148+5251 &  11^{\rm{h}}48^{\rm{m}}16^{\rm{s}}.65 &  +52\degr51\arcmin50\arcsec.39 &  6.4189\pm0.0006 &  [\ion{C}{2}] & 11 & -27.62 & 4.70\pm0.03& 1.16\pm0.01\\
J1250+3130 &  12^{\rm{h}}50^{\rm{m}}51^{\rm{s}}.93 &  +31\degr30\arcmin21\arcsec.90 &  6.138\pm0.005\tablenotemark{*} &  \ion{Mg}{2} & 15 &  -26.53 & 4.91\pm0.29& 2.17\pm0.13\\
J1306+0356 & 13^{\rm{h}}06^{\rm{m}}08^{\rm{s}}.27 &  +03\degr56\arcmin26\arcsec.36 &  6.0337\pm0.0004\tablenotemark{*} &  [\ion{C}{2}] & 14 &  -26.81 & 6.51\pm0.02& 2.48\pm0.01\\
J1319+0950 &  13^{\rm{h}}19^{\rm{m}}11^{\rm{s}}.30 &  +09\degr50\arcmin51\arcsec.52 &  6.1330\pm0.0007 &  [\ion{C}{2}] & 7 & -27.05 & 4.99\pm0.04& 1.68\pm0.01\\
J1335+3533 &  13^{\rm{h}}35^{\rm{m}}50^{\rm{s}}.81 &  +35\degr33\arcmin15\arcsec.82 &  5.9012\pm0.0019 &  CO & 4 & -26.67 & 0.70\pm0.10& 0.29\pm0.04\\
J1411+1217 &  14^{\rm{h}}11^{\rm{m}}11^{\rm{s}}.29 &  +12\degr17\arcmin37\arcsec.28 &  5.904\pm0.002 &  \ion{Mg}{2} & 2 & -26.69 & 4.61\pm0.13& 1.88\pm0.05\\
J1602+4228 &  16^{\rm{h}}02^{\rm{m}}53^{\rm{s}}.98 &  +42\degr28\arcmin24\arcsec.94 &  6.083\pm0.005\tablenotemark{*} &  \ion{Mg}{2}  & 15 &  -26.94 & 6.82\pm0.29& 2.43\pm0.10\\
J1623+3112 &  16^{\rm{h}}23^{\rm{m}}31^{\rm{s}}.81 &  +31\degr12\arcmin00\arcsec.53 &  6.2572\pm0.0024 &  [\ion{C}{2}] & 12 &  -26.55 & 5.05\pm0.14& 2.21\pm0.06\\
J1630+4012 &  16^{\rm{h}}30^{\rm{m}}33^{\rm{s}}.90 &  +40\degr12\arcmin09\arcsec.69 &  6.065\pm0.007 &  \ion{Mg}{2}  & 3 & -26.19 & 5.25\pm1.03& 2.79\pm0.55\\
J1641+3755 &  16^{\rm{h}}41^{\rm{m}}21^{\rm{s}}.73 &  +37\degr55\arcmin20\arcsec.15 &  6.047\pm0.003 &  \ion{Mg}{2} & 5 & -25.67 & 4.00\pm0.18& 2.80\pm0.13\\
J2054$-$0005 &  20^{\rm{h}}54^{\rm{m}}06^{\rm{s}}.49 &  -00\degr05\arcmin14\arcsec.80 &  6.0391\pm0.0001 &  [\ion{C}{2}] & 7 & -26.21 & 3.12\pm0.01& 1.64\pm0.01\\
J2229+1457 &  22^{\rm{h}}29^{\rm{m}}01^{\rm{s}}.65 &  +14\degr57\arcmin09\arcsec.00 &  6.1517\pm0.0005 &  [\ion{C}{2}]  & 11 &  -24.78 & 0.48\pm0.04& 0.54\pm0.04\\
J2329$-$0301 &  23^{\rm{h}}29^{\rm{m}}08^{\rm{s}}.28 &  -03\degr01\arcmin58\arcsec.80 &  6.4164\pm0.0008\tablenotemark{*} &  [\ion{C}{2}] &13&   -25.25 & 2.73\pm0.04& 2.39\pm0.04\\
    \enddata
    \tablenotetext{*}{The redshifts updated from \citet{Eilers2017}\citep{Willott2017,Decarli2018,Shen2019}}
    \tablecomments{Same as Table \ref{tab:sample}, but for the bright sample from \citet{Eilers2017}.
    Absolute magnitudes $M_{1450}$ are taken from \citet{Banados2016}.\\
    References for redshifts. (1)\citet{Jiang2007}, (2)\citet{Kurk2007}, (3)\citet{Carilli2010}, (4)\citet{Wang2010}, (5)\citet{Willott2010}, (6)\citet{DeRosa2011}, (7)\citet{Wang2013}, (8)\citet{Willott2013}, (9)\citet{Banados2015}, (10)\citet{Becker2015}, (11)\citet{Willott2015}, (12)\citet{Eilers2017}, (13)\citet{Willott2017},(14)\citet{Decarli2018}, (15)\citet{Shen2019}
    }
    
\end{deluxetable*}
 
\begin{figure}[htb]
     \centering
     \includegraphics[width=\linewidth]{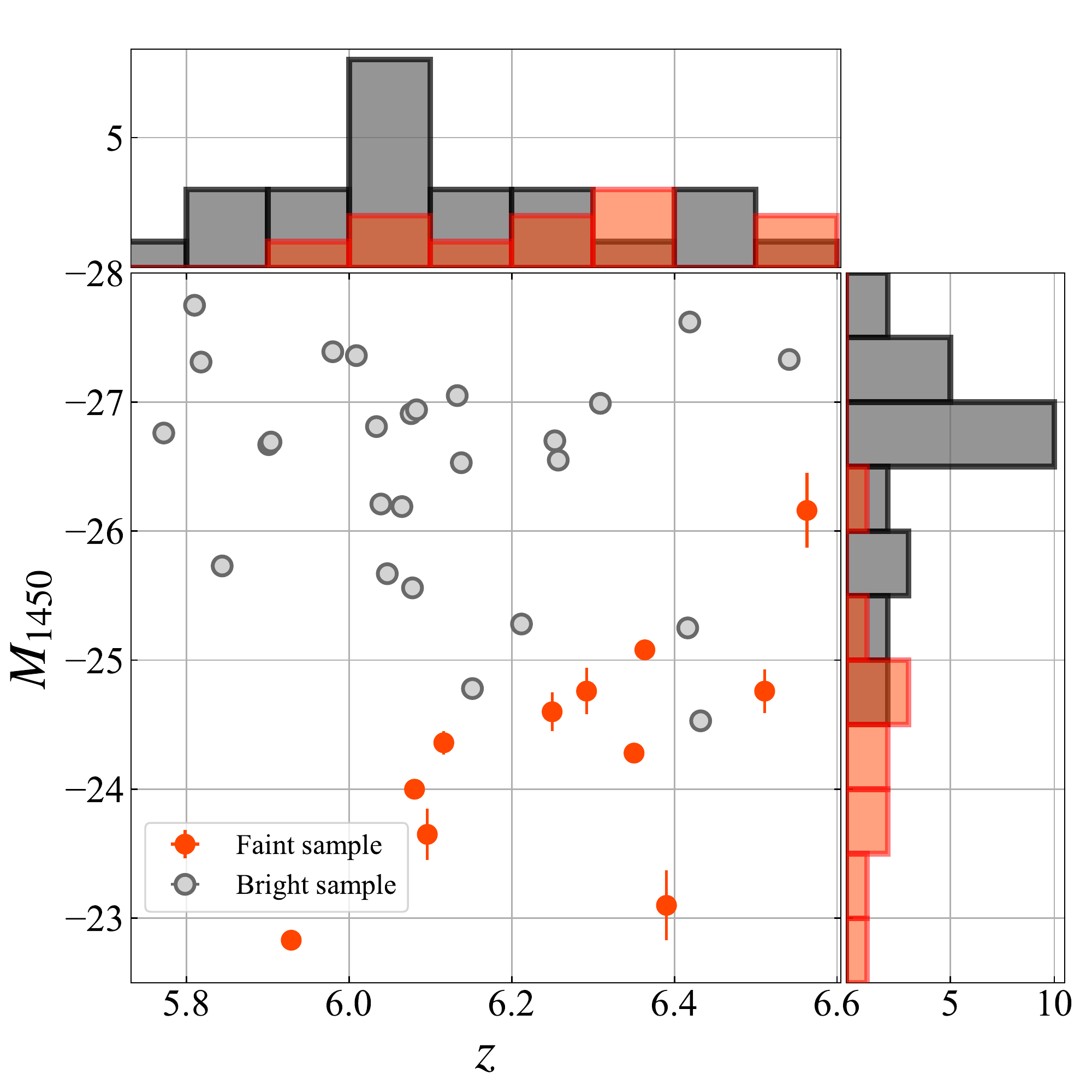}
     \caption{
     The redshift and $M_{1450}$ distribution of our sample.
     The orange circles show the faint quasars, and the grey circles show the brighter quasars. 
     The magnitude errors are not given in \citet{Eilers2017}, but are likely to be small.}
     \label{fig:mag}
 \end{figure}

 Figure \ref{fig:mag} compares magnitudes and redshifts between our new sample and that of \citet{Eilers2017}.
 Our new sample is 2-3 mag fainter than that of \citet{Eilers2017}.
 The combined sample gives us a dynamic range of 5 magnitudes in luminosity.
 
\section{The Proximity Zone Size Measurements}\label{sec:method}
\subsection{Quasar Continuum Normalization}
We estimate the quasar intrinsic spectra after normalizing at rest 1280 \AA\ with principal component spectra (PCS) from a principal component analysis (PCA) of low-redshift quasar spectra.
 This approach is justified by lack of spectral evolution of quasars \citep[e.g.,][]{Jiang2009}.
In PCA,  the quasar spectrum, $q_i(\lambda)$, is  modeled as a mean quasar spectrum, $\mu(\lambda)$, and  a linear combination of PCS:
\begin{equation}
    q_i(\lambda)\sim\mu(\lambda)+\sum_{j=1}^m c_{ij}\xi_j(\lambda),
\end{equation}
where $i$ refers to a  $i$th quasar, $\xi_j(\lambda)$ is the $j$th PCS, and $c_{ij}$ is the weight.
We use the PCS from \citet{Suzuki2005}.
First, $c_{ij}'$, the weights for the spectrum redward of 1216\AA, are derived by
\begin{equation}
    c_{ij}' = \int_{1216{\mathrm{\mathring{A}}}}^{\lambda_{\rm upper}}[q_i(\lambda)-\mu(\lambda)]\xi_j(\lambda)d\lambda,
\end{equation}
 where $\lambda_{\rm upper}$ is the upper limit of available wavelength in each observed quasar spectrum.
\citet{Suzuki2005} produced PCS for 1216\ \AA\ to 1600\ \AA, while our faint sample usually has coverages up to $\sim1350$ \AA.

Then we use the projection matrix $\bm{X}$ to calculate $c_{ij}$, the weights for the whole intrinsic  spectrum, covering the entire spectral region between 1020\AA\ and 1600\AA, using
\begin{equation}
    c_{ij}= c_{ij}'\cdot\bm{X}.
\end{equation}
 The projection matrix $\bm{X}$ is also taken from \citet{Suzuki2005}.
It is the matrix which satisfies the relation $\bm{C}=\bm{D}\cdot\bm{X}$, where $\bm{C}$ and $\bm{D}$ are the weights of principal components of the whole and the redward of quasar spectrum derived in \citet{Suzuki2005}, respectively.

\citet{Eilers2017} mainly used PCS from \citet{Paris2011}, but we use PCS and the projection matrix from \citet{Suzuki2005},  who constructed the PCS using fainter quasars at $z<1$ than those of \citet{Paris2011}.
However, we found no significant difference in the  results between the two.
\deleted{ Following \citet{Eilers2017}, }We  use five PCS for all quasar spectra.
\begin{figure*}[p]
     \centering
     \includegraphics[width=\linewidth]{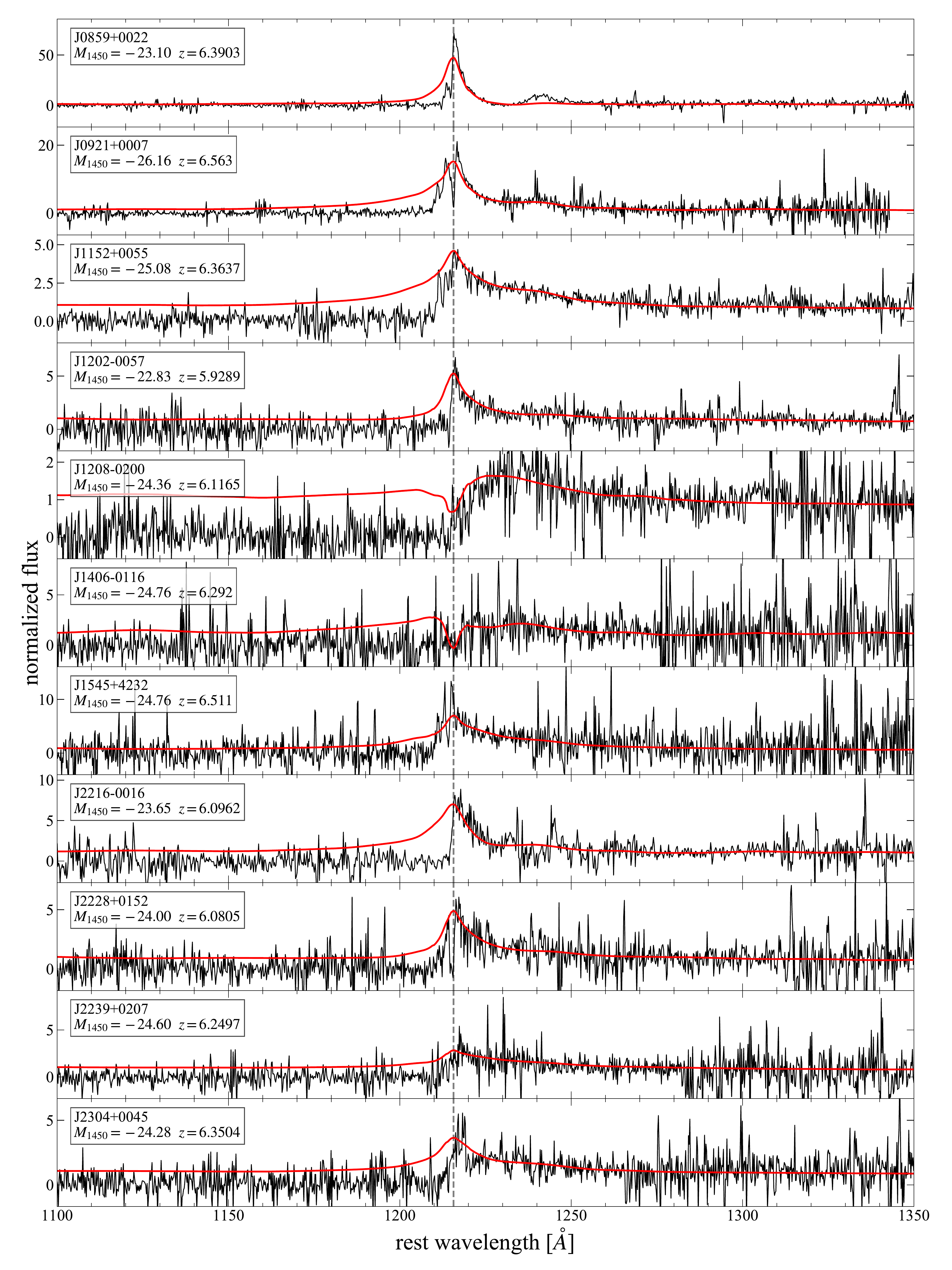}
     \caption{Predicted  intrinsic spectra using principal component analysis (PCA). The black and red curves show  the observed and  the predicted spectrum of each quasar, respectively. The dashed vertical line indicates $1215.67$\AA.}
     \label{fig:pca}
 \end{figure*}
 The spectra of our faint sample  and the PCA fits are shown in Figure \ref{fig:pca}.

\subsection{Measuring Proximity Zone Sizes}\label{sec:measureRp}
\begin{figure*}[ht]
     \centering
     \includegraphics[width=\linewidth]{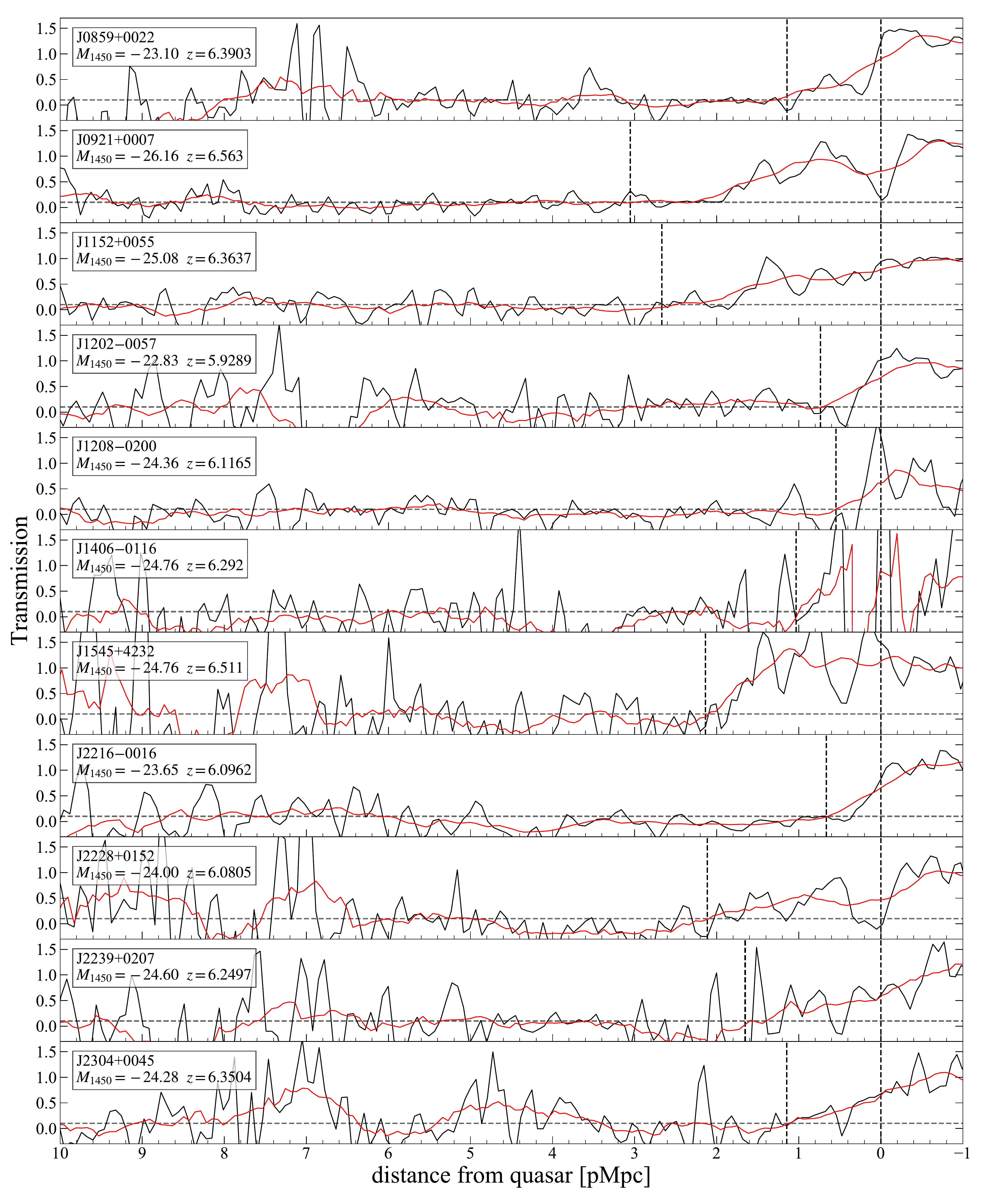}
     \caption{ Transmission spectra of the faint sample.
     The black and red curves  show the quasar spectra smoothed with  two pixels boxcar and a  resolution of 20 \AA, respectively.
     The horizontal dashed lines represent a flux level of 10\%. The vertical dashed lines show the points where normalized flux first  drops below the 10\% (left) and the quasar redshift (right).}
     \label{fig:nz}
 \end{figure*}
We adopt the same definition of proximity zone size as  was used in \citet{Fan2005}.
It is the physical distance between the quasar host galaxy ($z_{\rm{Q}}$) and  the point where the transmitted flux ratio first drops below 0.1, 
 using the observed quasar spectrum after smoothing to a resolution of 20\ \AA\ in the observed frame ($z_{\rm{GP}}$).
We regard the first of three consecutive pixels blueward of Ly$\alpha$ as the end of  the proximity zone  \citep{Eilers2017}, and calculate proximity zone sizes using
\begin{equation}
    R_p = \frac{D_{\rm{Q}}-D_{\rm{GP}}}{1+z_{\rm{Q}}},
\end{equation}
where $D_{\rm{Q}}$ and $ D_{\rm{GP}}$ are the comoving distances implied by $z_{\rm{Q}}$ and $z_{\rm{GP}}$, respectively.
Figure \ref{fig:nz} shows the  continuum-normalized spectra around the proximity zone of each quasar  in our faint quasar sample,
 and the measured $R_p$ are listed in Tables \ref{tab:sample} and \ref{tab:Esample}.
 
\deleted{Considering the errors on $R_p$, \citet{Fan2005} estimated that smoothing at 20 \AA\ resolution yields $\Delta R_p \sim0.6$ pMpc.
We add this error in quadrature to the errors due to  the redshift uncertainty.
We note that  for our faint quasars,  the smoothing error  dominates the uncertainties in $R_p$ because the redshifts of our sample are all accurately measured by [\ion{C}{2}] or \ion{Mg}{2}.}

In general, the observed wavelength range of the rest-UV spectrum of $z\sim6$ quasars is limited, which could cause a poor prediction of the intrinsic spectrum using PCA.  
The NIR spectra are available for five quasars, J0859$+$0022, J1152$+$0055, J1208$-$0200, J2216$-$0016, and J2239+0207 \citep{Onoue2019}, which extend a spectral coverage much further to $\sim 2.5\ \micron$.
The NIR spectra of the former two, J0859+0022 and J1152+0055, were taken by the Very Large Telescope/X-SHOOTER, and the latter three were taken by the Gemini Near-InfraRed Spectrograph (GNIRS). 
In addition, for former two, the optical spectra taken by X-SHOOTER are available, which have higher resolutions and deeper integrations than the FOCAS/OSIRIS one \citep{Onoue2019}.
\edit1{We use their optical and NIR spectra to measure $R_p$ of these five quasars.
When we use only the optical spectra, the resultant $R_p$ are $1.22\pm0.06$ pMpc, $2.60\pm0.03$ pMpc, $0.62\pm0.01$ pMpc, $0.66\pm0.02$ pMpc, and $1.31\pm0.02$ pMpc.}
\edit1{Three of them, J0859+0022, J1208$-$0200, and J2216$-$0016 are consistent within the errors with those in Table \ref{tab:sample}, suggesting that our $R_p$ measurements are not
significantly affected by the limited wavelength coverage.
}

Several quasars show a weak or no Ly$\alpha$ emission line, i.e., J1208$-$0200 and J1406$-$0116, as is often seen in $z\sim6$ quasars \citep[e.g.,][]{Banados2014}.
This could  give rise to a poor PCA fit around the wavelength of Ly$\alpha$ showing apparent negative Ly$\alpha$ emission.
As a test, we remeasured $R_p$ for these \edit1{two} quasars using a simple power-law  fit to the continuum \citep{Fan2005,Carilli2010} over wavelength intervals devoid
of emission lines at 1275--1295 and 1325--1335 \AA\  in the rest frame.
The resultant $R_p$ are \edit1{$0.55\pm0.01$ pMpc and $1.03\pm0.41$ pMpc }for J1208$-$0200 and J1406$-$0116, respectively.
The $R_p$ of J1406$-$0116 is larger than the PCA measurement, which gives extremely small $R_p$, probably due to its relatively noisy spectrum.
\edit1{Although it is hard to determine which measurement is likely to be more accurate for J1406$-$0116 due to their relatively poor quality spectra, we decide to adopt PCA measurement to keep consistency with other sample.}
The $R_p$ of J1406$-$0116 might have large uncertainty, but we find this discrepancy does not affect the final result of  the luminosity (Sec.\ref{sec:M-Rp}) and redshift (Sec.\ref{sec:z-Rpcorr}) dependences.

\begin{figure}[htb]
    \centering
    \includegraphics[width=\linewidth]{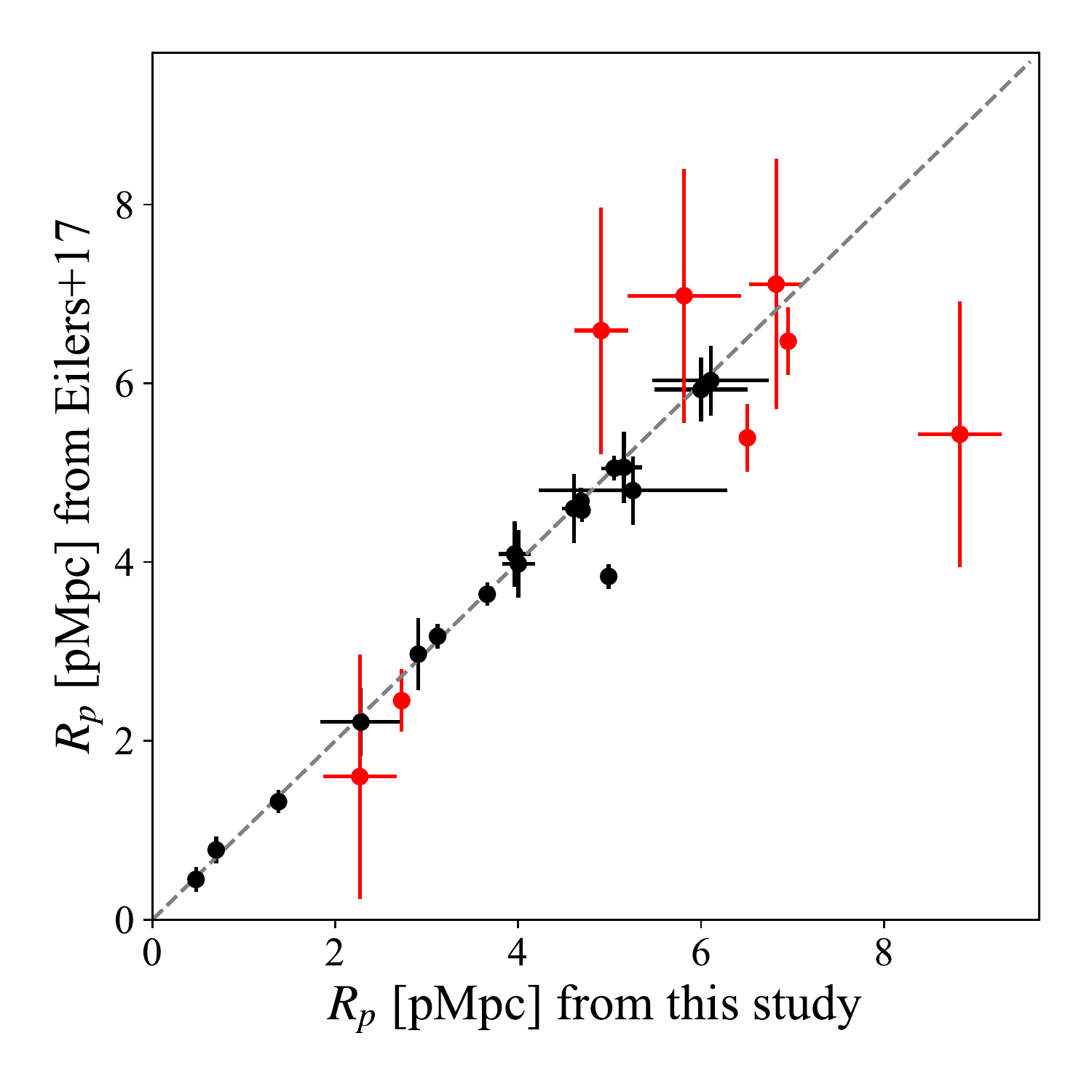}
    \caption{Comparison between $R_p$ mesured in this study and in \citet{Eilers2017}.
    The quasars whose redshifts are updated from \citet{Eilers2017} are shown  as red points.
     \deleted{We note that \citet{Eilers2017} did not consider the smoothing error.}
    }
    \label{fig:Rp-Rp}
\end{figure}

Figure \ref{fig:Rp-Rp} shows a comparison in $R_p$ for the bright sample between our measurement and \citet{Eilers2017}. They are in good agreement  with each other except for those quasars whose redshifts have been updated using data from \edit1{\citet{Willott2017},} \citet{Decarli2018}, and \citet{Shen2019}.
\edit1{Consequently, this moderately change the PCA fit, supporting our previous statement that accurate redshift measurements are needed for the accurate $R_p$ measurements.}

\edit1{It should be noted that the spectrum signal-to-noise ratios of the faint sample are generally lower than those of the bright sample.
We check the $R_p$ uncertainties due to the spectral noise by the following Monte Carlo simulation using the noise spectra.
In this process, the flux of each spectral pixel was associated with a random error perturbed within the measured $1\sigma$ error.
We generated 100 mock spectra and repeated the PCA analysis.
The $R_p$ uncertainty of the trials is found to be $0.33\pm0.32$ pMpc, which is comparable to the errors due to the redshift uncertainty, except for the two quasars, J0921+0007 (0.71 pMpc) and J1202$-$0057 (1.06 pMpc). 
We also found the error is almost negligible for the two quasar spectra, J0859+0022 and J1152+0055, taken by X-SHOOTER.
The 16th and 84th percentiles of these uncertainties are 0.07 pMpc and 0.64 pMpc, respectively.
We confirm this additional error do not significantly change the result.
It is not clear how large the error for the bright sample from \citet{Eilers2017}.
To make a fair comparison with the bright sample, this error is not taken into account. }

\section{Results and Discussion}\label{sec:results}
\subsection{Proximity Zone Sizes Using Stacked Spectra}
 To illustrate the luminosity dependence, we create mean-stacked spectra of the faint and the bright samples, and measure $R_p$ for both.
These spectra were generated by normalizing each  spectrum by  the flux density at 1280 \AA\  of  PCA fit,
converting to the rest frame, and  then mean-stacking.
When we measure $R_p$ of these spectra, we assume the mean redshifts of each sample as  the stacked quasar redshift.
\begin{figure*}[ht]
     \centering
     \includegraphics[width=\linewidth]{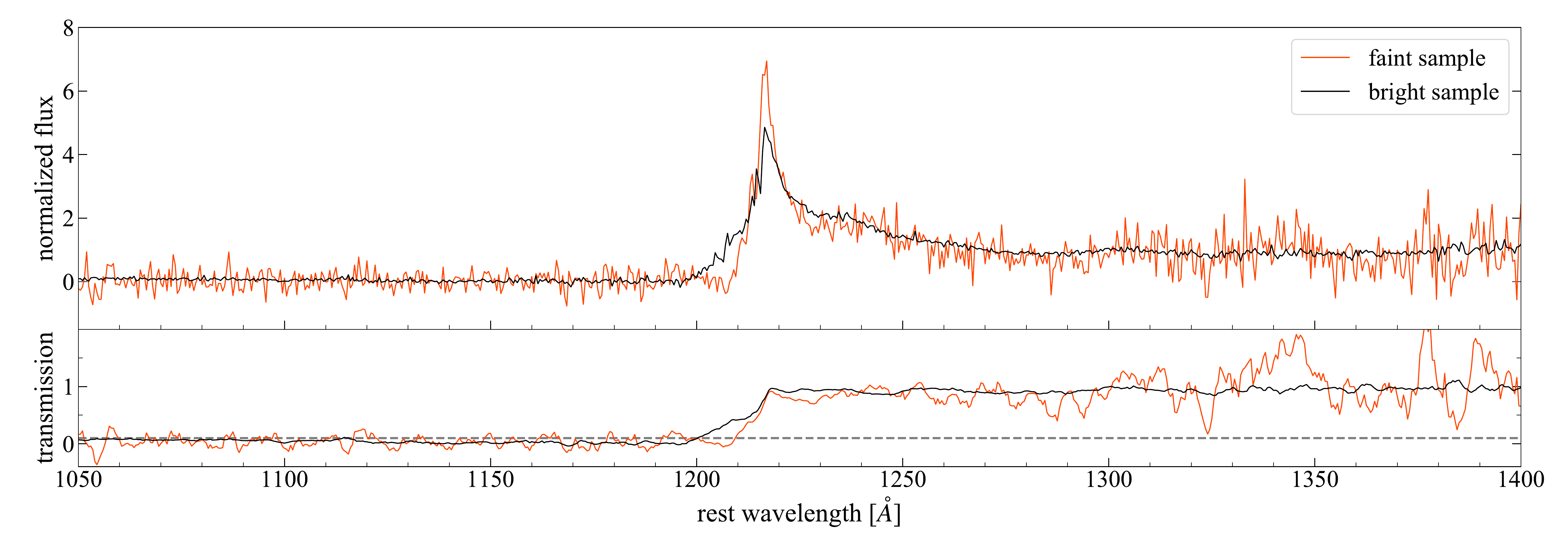}
     \caption{(top) Rest-frame composite spectra of  the faint sample (red) and the bright sample (black).
     (bottom) Transmission spectra smoothed to a resolution of 20 \AA\ in observed  wavelengths.
     The grey dashed line shows a flux level of 10\%.}
     \label{fig:stack}
\end{figure*}
Figure \ref{fig:stack} shows the  two stacked spectra.
The $R_p$ of our faint and bright quasar sample are \edit1{$R_p=2.23\pm0.03$ pMpc and $5.45\pm0.06$ pMpc}, respectively.
Our faint sample shows significantly smaller $R_p$ than that of the bright sample  as Equation (\ref{eq:Rp_haiman}) predicts.

 \citet{Matsuoka2019} suggested the faint sample shows systematically narrower Ly$\alpha$ emission.
The composite spectra shown in Figure \ref{fig:stack} based on more accurate systemic redshift definitely  shows that the faint quasar sample has narrower Ly$\alpha$ emission than  the brighter sample. 
The reason  for this is unclear, it may be partly due to contamination from narrow line quasars with exceptionally narrow Ly$\alpha$ emission lines \citep{Kashikawa2015,Matsuoka2019}.  

\subsection{Luminosity Dependence}
\label{sec:M-Rp}

\begin{figure*}[htb!]
     \centering
     \includegraphics[width=0.8\linewidth]{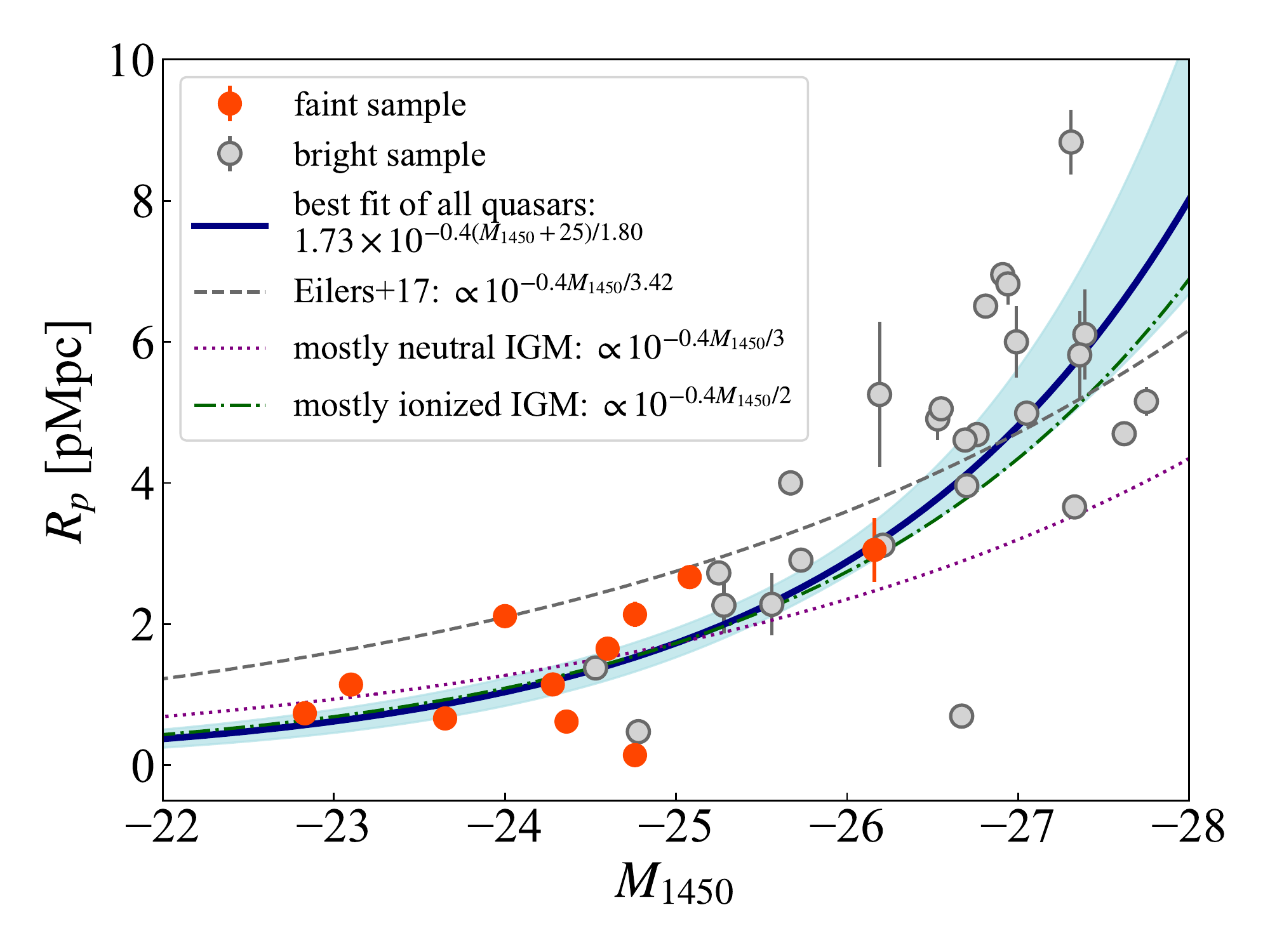}
     \caption{Proximity zone sizes $R_p$ as a function of  the quasar magnitudes $M_{1450}$. 
     The orange and grey circles indicate $R_p$  for the faint and the bright sample, respectively.
     The blue solid line  and the shaded region show the best  power-law fit to the  measurements and its 1$\sigma$ uncertainty from bootstrapping errors, respectively. 
     The grey dashed line shows the best fit in \citet{Eilers2017}.
     The purple dotted and green dot-dashed curves show the theoretical predictions assuming  a mostly neutral IGM (Eq.\ref{eq:Rp_haiman}) and  a mostly ionized IGM \citep{Bolton2007}, respectively.
     }
     \label{fig:M-Rp}
 \end{figure*}
Figure \ref{fig:M-Rp} shows the relation between quasar proximity zone sizes $R_p$ and quasar absolute magnitude $M_{1450}$.
 We define $\alpha$ as a power-law index of luminosity dependence of $R_p$ ($R_p\propto10^{-0.4M_{1450}/\alpha}$).
A power-law  fit to our  measurements weighted by errors gives \edit1{$\alpha=1.80\pm0.29$;
\begin{equation}\label{eq:fit_lumi}
    R_p=(1.73\pm0.21)\times10^{-0.4\times(M_{1450}+25)/(1.80\pm0.29)}\ \mathrm{pMpc}.
\end{equation}}
In the fit, we weight the measurement by the errors.
The 1$\sigma$ uncertainty of this fit is calculated by bootstrapping the fit 1000 times.
 The best fit described in \citet{Eilers2017} is
\begin{equation}
    R_p=4.71\times10^{-0.4\times(M_{1450}+27)/3.42}\ \mathrm{pMpc}.
\end{equation}
We normalize the relation at $M_{1450}=-25$, which is the  mid-point of our data, while \citet{Eilers2017}  normalized at $M_{1450}=-27$. 
We obtain  a steeper relation than the best fit in \citet{Eilers2017}.
The luminosity dependence of proximity zone sizes could, in principle, depend on the IGM ionization state. 
Equation (\ref{eq:Rp_haiman}) indicates that $R_p$  is proportional to $\alpha=3$ in the case of  a neutral IGM, while \citet{Bolton2007}  showed analytically that the proximity zone
size scales as $\alpha=2$, in the case of  a highly ionized IGM.
Our result on the luminosity dependence is close to the prediction for the ionized IGM, suggesting that most of  the surrounding IGM is ionized at $z\sim6$.
 As described in the introduction, the observed proximity zone sizes $R_p$  are not strictly identical to the  distances to the ionization front, and actual luminosity dependence could be affected by the detailed ionizing process; therefore, radiative transfer simulations would be required to  make further comparison with our result.
 The simulation of \citet{Eilers2017}, whose fit over the luminosity range of their sample, predicts  a scaling  of $\alpha=2.35$ in a highly ionized IGM. 

\subsection{Redshift Evolution}
\label{sec:z-Rpcorr}
\begin{figure*}[ht]
     \centering
     \includegraphics[width=0.8\linewidth]{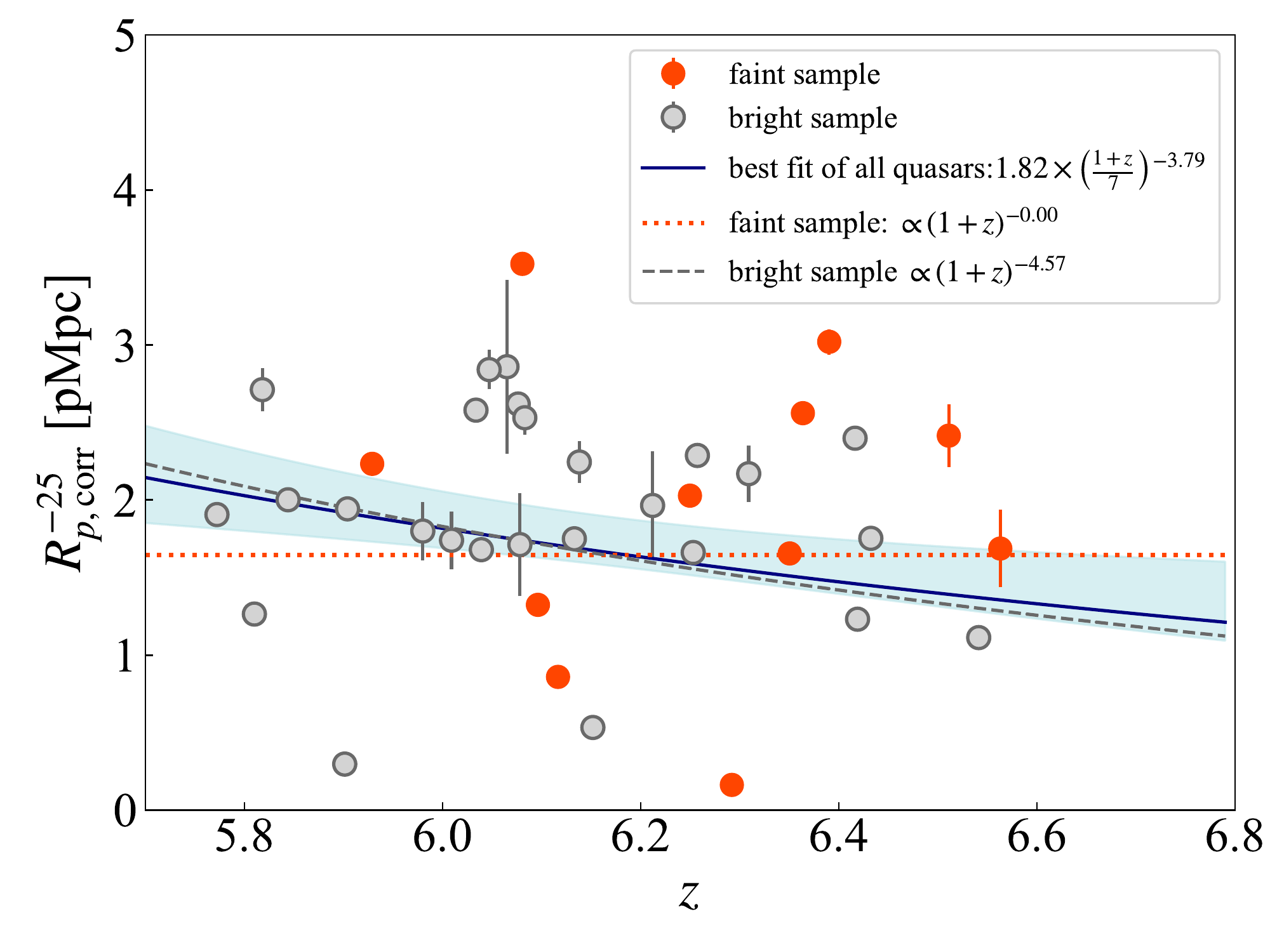}
     \caption{Redshift evolution of rescaled proximity zone sizes $R_{p, \rm corr}^{-25}$.
     The orange and grey circles indicate $R_{p, \rm corr}^{-25}$  values for the faint and the bright sample, respectively.
     The blue solid line shows the best power-raw fit to the measurement with 1$\sigma$ uncertainty from bootstrapping errors. 
     The orange dotted line and grey dashed line show the best fit to the measurement of the faint and the bright sample, respectively.}
     \label{fig:z-Rp}
\end{figure*}

In order to examine  the redshift evolution of $R_p$, 
we use  the luminosity scaling of our data from Equation (\ref{eq:fit_lumi});
\begin{equation}
    R_{p, \rm corr}^{-25}\footnote{\edit1{We denote our luminosity-corrected $R_p$ measurement as $R_{p,\rm corr}^{-25}$, normalized at $M_{1450}=-25$, while \citet{Eilers2017} nomalized at $M_{1450}=-27$.}}=R_p\times10^{0.4\times(M_{1450}+25)/\edit1{1.80}}
    \label{eq:Rpcorrection}
\end{equation}
to rescale the measured $R_p$, and the derived $R_{p,\rm corr}^{-25}$ are listed in Tables \ref{tab:sample} and \ref{tab:Esample}.
Figure \ref{fig:z-Rp} shows the redshift evolution of proximity zone sizes corrected by absolute magnitude.
 We define $\beta$ as a power-law index of a dependence on redshifts of $R_{p,\rm corr}^{-25}$ ($R_{p,\rm corr}^{-25}\propto(1+z)^\beta).$
A power-law fit using both the faint and the bright quasar sample gives \edit1{$\beta=-3.79\pm1.72$;
\begin{equation}\label{eq:z-Rpcorr_all}
    R_{p, \rm corr}^{-25}=(1.82\pm0.18)\times\left(\frac{1+z}{7}\right)^{-3.79\pm1.72} {\rm pMpc}.
\end{equation}}
The 1$\sigma$ uncertainty is calculated by bootstrapping.
The redshift dependency is \deleted{slightly} steeper than the best fit by \citet{Eilers2017}, $\beta=-1.44$.
\edit1{When we do not weight the measurement by the errors as \citet{Eilers2017} did not, a power-law fit gives $\beta=-1.43$, consistent with \citet{Eilers2017}.}
\deleted{It lies between the predictions of the ionized ($\beta=-2.45$) and neutral ($\beta=-1.62$) radiative transfer simulations of \citet{Eilers2017}. }
It is substantially shallower than that found in earlier studies, 
which presented  a linear fit to their corrected measurements  for $z>5.7$ quasars \citep{Carilli2010,Venemans2015}.
 When we correct $R_p$ using Equation (\ref{eq:Rpcorrection}) for measurements of \citet{Carilli2010} and \citet{Venemans2015},
the power-law fit gives \edit1{$\beta=-8.40\pm0.91,\ -7.83\pm0.36$}, respectively.
Thus we conclude that our $R_p$  shows a mild evolution at $z\sim6$.

When using only  the bright quasar sample, we obtain 
\edit1{
\begin{equation}
    R_{p, \rm corr}^{-25}=(1.78\pm0.19)\times\left(\frac{1+z}{7}\right)^{-4.86\pm1.94} {\rm pMpc},
\end{equation}}
as the best fit, which is \deleted{perfectly} consistent with Equation (\ref{eq:z-Rpcorr_all}) for the  full sample.
 We note that \citet{Eilers2017}  corrected the luminosity dependence of $R_p$ with factor $\alpha=2.35$ rather than their best fit result, $\alpha=3.42$. 
On the other hand,  the best fit using the faint sample only is,\edit1{
\begin{equation}
    R_{p,\rm corr}^{-25}=(1.68\pm0.44)\times\left(\frac{1+z}{7}\right)^{-0.00\pm2.00} {\rm pMpc}.
\end{equation}}
No redshift dependence is detected, and the $R_{p,\rm corr}^{-25}$ is slightly smaller than that of the bright sample only.
All these three fits show shallow  to no redshift evolution.
 A Kolmogorov-Smirnov test to access the statistical significance of the difference between the faint sample and the bright sample, 
yielded \edit1{$p=0.93$}, suggesting that the difference  between the two is statistically not significant;
 the two samples have almost the same distribution of corrected proximity zone  sizes, albeit the faint sample has large errors. 
The size of our faint sample is still small, and  we will be able to make firmer conclusions as the sample of faint quasars with accurate redshifts grows.

\subsection{Young Quasar Candidates with Exceptionally Small Proximity Zones}\label{sec:discussion}
\deleted{In this study, we  have measured the proximity zone sizes of ten  faint ($M_{1450}>-26.16$) and 26 bright quasars at $z\sim6$.
We  have confirmed that the redshift evolution is shallow, consistent with \citet{Eilers2017} and previous simulation results \citep{Bolton2007,Davies2019}.}
\citet{Davies2019} presented radiative transfer simulation to investigate the behavior of $R_p$.
 They found that the only quasars with $R_{p, {\rm corr}}\lesssim2.5$ pMpc are young ($t_Q\lesssim10^4$yr),  where $R_p$ is normalized to an absolute magnitude of $M_{1450}=-27$.
This corresponds to $R_{p,{\rm corr}}^{-25}\lesssim\edit1{0.90}$ pMpc with our normalization at $M_{1450}=-25$ using Equation (\ref{eq:Rpcorrection}).
There are \edit1{two} quasars that meet this criterion in the faint sample, J1208$-$0200 and J1406$-$0116, and two in the bright sample, J1335$+$3533 and J2229$+$1457, which  \citet{Eilers2017} also suggested to have an exceptionally small proximity zone size.
\edit1{We should note that $R_p$ measurement of J1406$-$0116 might be inaccurate due to poor PCA fit (see Sec.\ref{sec:measureRp}).}
These \edit1{four} quasars may be young,
but it is also possible that neutral gas lying along the quasar sightline truncates the proximity zones.
One  piece of evidence for such a clump of high column density neutral gas, such as Damped Ly$\alpha$ Systems (DLAs) and Lyman Limit Systems (LLSs), would be the presence of associated metal-line absorbers \citep{Eilers2017}.
\citet{Eilers2018} conducted spectroscopic  observations of J1335$+$3533 and ruled out the possibility that its small $R_p$ is due to  an associated absorption system.
\deleted{However, the spectra of our faint quasar sample have insufficient S/N to identify very weak metal absorptions features. }
\edit1{J1208$-$0200 shows significant  absorption redward of Ly$\alpha$ emission line, implying the presence of a strong foreground absorption feature such as a proximate DLA, which could exhibit low-ionization metal absorption lines.
We search corresponding low-ionization metal absorption lines in the spectrum of J1208$-$0200, \ion{Si}{2} ($1260.42$ \AA\ and $1304.37 $ \AA), \ion{O}{1} ($1302.16$ \AA), and [\ion{C}{2}] ($1334.53$ \AA), and find no clear absorption features.
However, the spectra of our faint quasar sample, in general, have insufficient S/N to identify very weak metal absorption features. }

\begin{figure*}[htb!]
     \centering
     \includegraphics[width=\linewidth]{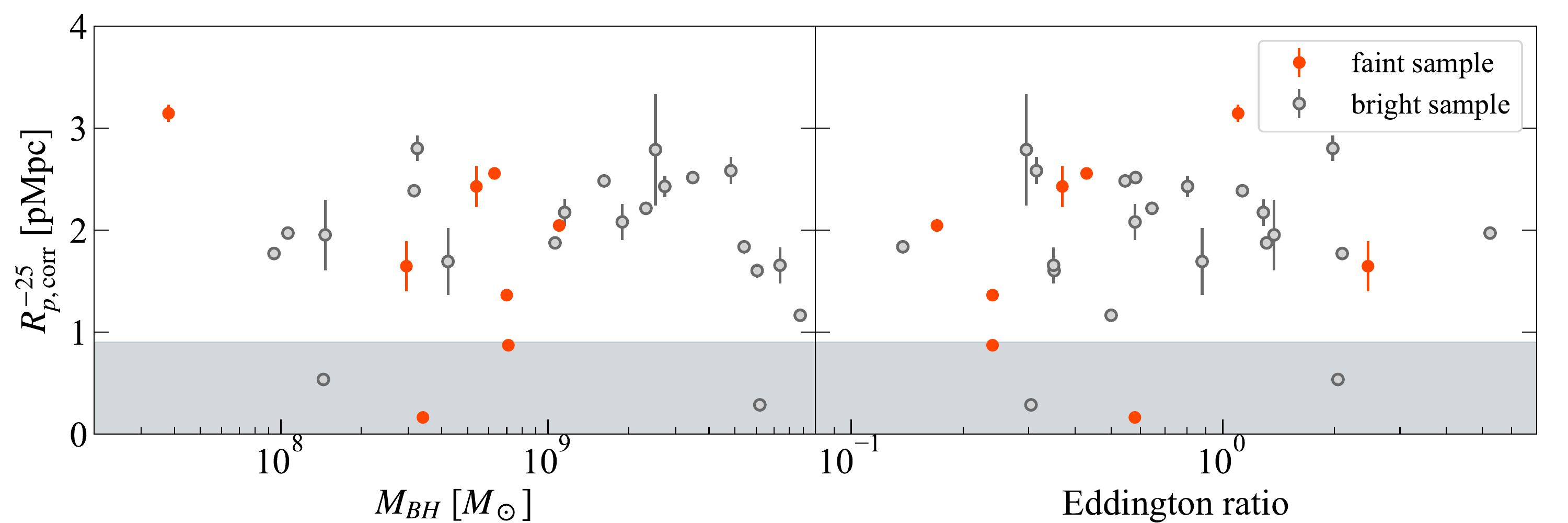}
     \caption{The correlation between $R_{p,\rm corr}$ and black hole mass (left) and Eddington ratio (right). 
     The \ion{Mg}{2}-based mass measurements have systematic uncertainties of 0.5 dex \citep{Shen2013}.
     The orange and grey dots show the faint and bright sample, respectively. 
     The shaded region indicates $R_{p,{\rm corr}}^{-25}\leq\edit1{0.90}$ pMpc. }
     \label{fig:BHmass_Edd}
\end{figure*}

Eight objects in our faint sample have \ion{Mg}{2}-based measurements of black hole mass $M_{\rm BH}$ and Eddington ratio \citep[][Onoue et al. in prep.]{Onoue2019}, and as do for 21 objects in the bright sample \citep{Shen2019}.
We examine the correlation between black hole mass, Eddington ratio and proximity zone size in Figure \ref{fig:BHmass_Edd}.
Young quasar candidates suggested by extremely small $R_p$ tend to have smaller $M_{\rm BH}$ and lower Eddington ratio, though there is no clear correlation.

\citet{Meyer2019} found that the average blueshift of the \ion{C}{4} emission line with respect to low-ionization lines in quasar spectra increases significantly at $z>6$.
\edit1{The authors interpreted this trend} as due to a strong outflows, likely related to the relative youth of high-$z$ quasars.
Five objects in our faint sample, J0859+0022, J1152+0055, J1208$-$0200, J2216$-$0016 and J2239+0207, were found to have significant \ion{C}{4} blueshift with respect to \ion{Mg}{2} lines \citep[Table 3]{Onoue2019}.
Interestingly, the quasars with the most extreme blueshifted \ion{C}{4} \edit1{is J1208$-$0200 (1830 km s$^{-1}$), and J2216$-$0016 (1170 km s$^{-1}$) is the second among these five quasars. 
J1208$-$0200 is suggested as young quasar candidate because of its extreme small proximity zone, and J2216$-$0016 also exhibits relatively small proximity zone, $R_{p, \rm corr}^{-25}=1.36$ pMpc.}
Both observational quantities consistently \edit1{indicate} the young age of quasars.
\citet{Farina2019} recently conducted a sensitive search for extended Ly$\alpha$ halos around $z\sim6$ quasars with MUSE.
While they detected significant extended Ly$\alpha$ emissions around 12 quasars, one of the young quasar candidates, J2229+1457 does not show an extended Ly$\alpha$ halo.
Another young quasar candidate from our faint sample, J2216$-$0016 is also observed by \citet{Farina2019} and it has neither extended Ly$\alpha$ halo. 
As discussed in \citet{Farina2019}, a young quasar with $t_Q<10^4$ yr does not have enough time to light up an extended Ly$\alpha$ halo with more than 10 pkpc radius, large enough to be observed by their survey.
Along with the proximity zone size and the \ion{C}{4} blueshift, the Ly$\alpha$ halo extension could be another promising observational diagnostic of young quasars; however, a larger sample is obviously required to make a clear conclusion. 

\citet{Eilers2020} additionally found four young quasar candidates with extremely small proximity zone sizes. 
They constrained the fraction of young quasars within the luminous ($M_{\rm UV}\lesssim-25$) quasars as $5\%<f_{\rm young}<9\%$, while it is interesting to note that our faint sample exhibit a high fraction as \edit1{$\sim2/11\sim18\%$.}
Future observations of such first quasars will reveal nature of quasar activity, such as lifetime and duty cycle. 

\section{Summary}\label{sec:summary}
In this paper, we measure the proximity zone sizes for a sample which consists of \edit1{eleven} faint $z\sim6$ quasars discovered by the SHELLQs project, and 26 luminous $z\sim6$ quasars which were analyzed in \citet{Eilers2017}. 
Our faint sample significantly expands the dynamic range of quasar luminosity to examine more common and numerous quasar environments in the reionization era.
It is essential to use precise redshifts for accurate $R_p$ measurement.
All redshifts of our quasar sample have been accurately  measured  from the [\ion{C}{2}], \ion{Mg}{2}, or CO emission lines.

We estimate the intrinsic quasar spectra by PCA, 
using PCS from \citet{Suzuki2005}, 
and measure the size of the proximity zones.
The major results in this study are summarized below.
\begin{enumerate}
\item 
We compare the mean-stacked spectra based on the accurate systemic redshifts of our faint and bright samples.
The $R_p$ of the faint sample is significantly smaller than that of the bright sample.
The faint sample shows a narrower Ly$\alpha$ emission line than that of the bright sample.
\item 
The best fit of dependence of the proximity zone size on quasar luminosity is found to be $R_p\propto10^{-0.4M_{1450}/\edit1{1.80\pm0.29}}$.
This shallow relation is consistent with a theoretical model which assumes an ionized IGM \citep{Bolton2007}.
We use the best fit to rescale $R_p$  by quasar luminosity.
\item
Our results find a shallow redshift evolution, 
$R_{p, \rm corr}^{-25}\propto(1+z)^{\edit1{-3.79\pm1.72}}$.
This relation is \edit1{steeper than that of \citet{Eilers2017}, and significantly shallower than those of \citet{Carilli2010,Venemans2015}, all of which are based on the luminous quasar sample.}
The $R_p$ of the faint sample tend to be smaller than that of the bright sample, though with small significance. 
\item 
\edit1{Two} quasars in the faint sample and two in the bright sample show exceptionally small proximity zones ($R_{p,\rm corr}^{-25}<\edit1{0.90}$\ pMpc), implying that such quasars are young ($<10^4$ yr).
Some of these quasars have significantly blueshifted \ion{C}{4} emission lines and show no Ly$\alpha$ extended halos,
although statistical uncertainties still remain.
Further observation is required to uncover the local environment of high-$z$ quasars and the IGM state at the reionization epoch.
\end{enumerate}

\acknowledgments
{\edit1{We appreciate the anonymous referee
for helpful comments and suggestions that improved the manuscript.}

The Hyper Suprime-Cam (HSC) collaboration includes the astronomical communities of Japan and Taiwan, and Princeton University. 
The HSC instrumentation and software were developed by NAOJ, the Kavli Institute for the Physics and Mathematics of the Universe (Kavli IPMU), the University of Tokyo, the High Energy Accelerator Research Organization (KEK), the Academia Sinica Institute for Astronomy and Astrophysics in Taiwan (ASIAA), and Princeton University. 
Funding was contributed by the FIRST program from Japanese Cabinet Office, the Ministry of Education, Culture, Sports, Science and Technology (MEXT), the Japan Society for the Promotion of Science (JSPS), Japan Scienceprovide us and Technology Agency (JST), the Toray Science Foundation, NAOJ, Kavli IPMU, KEK, ASIAA, and Princeton University

This paper makes use of software developed for the Large Synoptic Survey Telescope (LSST). 
We thank the LSST Project for making their code available as free software at \url{http://dm.lsst.org}.

The Pan-STARRS1 Surveys (PS1) have been made possible through contributions of the Institute for Astronomy, the University of Hawaii, the Pan-STARRS Project Office, the Max-Planck Society and its participating institutes, the Max Planck Institute for Astronomy, Heidelberg and the Max Planck Institute for Extraterrestrial Physics, Garching, The Johns Hopkins University, Durham University, the University of Edinburgh, Queen's University Belfast, the Harvard-Smithsonian Center for Astrophysics, the Las Cumbres Observatory Global Telescope Network Incorporated, the National Central University of Taiwan, the Space Telescope Science Institute, the National Aeronautics and Space Administration under Grant No. NNX08AR22G issued through the Planetary Science Division of the NASA Science Mission Directorate, the National Science Foundation under Grant No. AST-1238877, the University of Maryland, and E\"{o}tv\"{o}s Lorand University (ELTE).
}

\facilities{Subaru, GTC, VLT:Kueyen, Gemini:Gillett}
\software{astropy\ \citep{AstropyCollaboration2013,AstropyCollaboration2018},
igmspec (\url{http://specdb.readthedocs.io/en/latest/igmspec.html})}

\bibliography{NZ_ref}

\end{document}